\documentclass[%
prl,
twocolumn,
amsmath,amssymb,
aps,
]{revtex4-2}

\usepackage{graphicx}% Include figure files
\usepackage{dcolumn}% Align table columns on decimal point
\usepackage{bm}% bold math
\usepackage[utf8]{inputenc}
\usepackage{todonotes}
\usepackage{soul}
\usepackage{todonotes}

\usepackage{lineno}%,hyperref}
\usepackage[colorlinks,linkcolor=red,citecolor=blue,urlcolor=blue]{hyperref}

\begin{document}

\title{Mechanical Optimization of Skateboard Pumping}

\author{Florian Kogelbauer}
\email{floriank@ethz.ch}
\affiliation{Department of Mechanical and Process Engineering, ETH Z\"{u}rich, Leonhardstrasse 27, 8092, Z\"{u}rich, Switzerland}
\author{Shinsuke Koyama}
\email{skoyama@ism.ac.jp}
\affiliation{The Institute of Statistical Mathematics, Tokyo 190-8562, Japan}
\author{Daniel E. Callan}
\email{dcallan@atr.jp}
\affiliation{Brain Information Communication Research Laboratory Group, ATR Institute International, Kyoto 619-0288, Japan}
\author{Shigeru Shinomoto}% https://orcid.org/0000-0002-5745-7736
\email{shigerushinomoto@gmail.com}
\affiliation{Brain Information Communication Research Laboratory Group, ATR Institute International, Kyoto 619-0288, Japan}
\affiliation{Graduate School of Informatics, Kyoto University, Kyoto 606-8501, Japan}
\affiliation{Research Organization of Science and Technology, Ritsumeikan University, Shiga 525-8577, Japan}

\date{\today}

\begin{abstract}
Skateboarders perform a reciprocating motion on a curved ramp, called pumping, by moving their bodies up and down perpendicular to the ramp surface. We propose a simple mechanical model for this pumping motion and solve the equation of motion explicitly in angular coordinates. This allows us to derive an optimal control strategy to maximize amplitude by dynamically adjusting the center of mass of the skateboarder. This optimal strategy is compared to experimental results for the motion of a skilled and an unskilled skateboarder in a half-pipe, validating that a skilled skateboarder follows the optimal control strategy more closely. 
\end{abstract}

\maketitle

%Reference for picture format: https://journals.aps.org/info/infoL.html

%\section{Intro}

Gaining popularity as an action sport in the early sixties, skateboarding grew into a social phenomenon that was even included as an official discipline in the 2021 Olympic Games \cite{hawks2019z}. From a scientific point of view, the mechanics of skateboarding have served as a versatile playground in classical dynamics and control theory \cite{varszegi2015skateboard, varszegi2016stabilizing}. Recently, skateboarding has also become popular as an appealing model problem from dynamic control in robotics \cite{kim2021bipedal,ito2020mechanism,iannitti2004minimum}. Mimicking the delicate motion of human bodies during various dynamically involved actions, state-of-the-art machine learning techniques are applied to capture performance data \cite{googleskate}. We also refer to 
\cite{kuleshov2010various,feng2014energy,hubbard1980human} for various in-depth considerations on the mechanics of the dynamics of skateboarding.

In this letter, we propose a different, bottom-up approach towards the optimal control of skateboarding by focusing on a particular confined motion with few degrees of freedom as a test case: pumping on a ramp. Despite its utmost simplicity, our minimal model is consistent with experimental data we obtained from comparing the performance of a skilled and an unskilled skateboarder. This approach not only avoids the computational complexity that goes along with high-dimensional, multi-degree-of-freedom models but is also in tune with reduced-order modeling and nonlinear normal modes \cite{kerschen2009nonlinear}. Indeed, the theory of nonlinear normal modes \cite{haller2016nonlinear}, i.e., the reduction to few effective degrees of freedom, is highly relevant in control problems \cite{peeters2009nonlinear} for mechanical systems. Apart from the appeal of a completely soluble, experimentally consistent example, our insights might prove useful as a basic test case for robotics as well \cite{syed2008motion}.
%\section{Model}

\begin{figure}[h]
\centering
\includegraphics[width=0.9\linewidth]{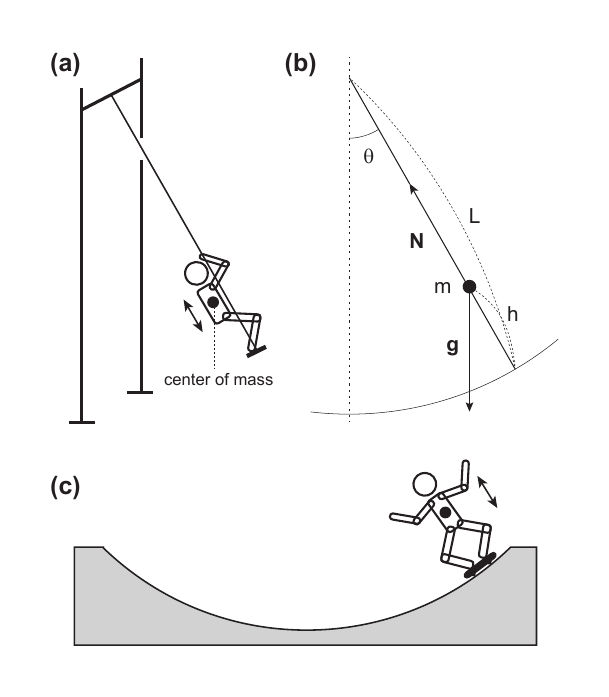}
\caption{\footnotesize{ Analogy of pumping motions. (a) Pumping on a swing; (b) The variable-length pendulum; (c) Pumping of a skateboarder on a cylindrical ramp. }}
\label{schematic}
\end{figure}

We aim to model the reciprocating, pumping motion of a skateboarder in the half-pipe. To this end, we propose a simple mechanical skateboarder model on a cylindrical ramp by analogy to the pumping motion on a swing, see Figure \ref{schematic}, which is itself classically modeled as a variable-length pendulum \cite{tea1968pumping, burns1970more}. The results obtained for the cylindrical ramp are then extended to the optimal control motion in the half-pipe by interpolating along the flat part of the ramp. The equation of motion for a frictionless variable-length pendulum takes the form ~\cite{BB13706363e, taylor2005classical}
\begin{equation}\label{maineq1}
(L-h) \ddot{\theta} - 2\dot{h}\dot{\theta}+g \sin \theta=0,
\end{equation}
where $\theta$ is the angle relative to a vertical axis, $L$ is the radius of the cylindrical ramp, $h$ is the height of the center of mass, and $g \sin \theta$ is the gravitational restoring force. A skateboarder controls his/her center of mass with respect to height, $h = h(t)$, perpendicular to the surface of the ramp. For more details on the dynamics of swinging, we refer to recent works on the pumping of a swing taking into account the detailed aspects of swinging specific to human body movement~\cite{klimina2020three, glendinning2020adaptive, koshkin2022swinging, hirata2023initial}. The mechanism in which a change in body height will result in a gain in rotational motion may be intuitively understood as a result of the conservation of angular momentum, particularly when the friction can be ignored. We therefore regard the height as an external control function to be optimised for maximal amplitude gain under certain constraints. Firstly, we assume that the height is constrained between a minimal bending and a maximal standing posture, $H_0\leq h\leq H_1$. Secondly, we assume that the energy input through the muscular activity of the human or the control force of a skateboarding robot is limited, leading to a constraint in the acceleration of the height $|\ddot{h}|\leq g$.

We consider equation \eqref{maineq1} together with kinetic friction by assuming that the wheels of the skateboard kinetic friction $\mu N$ at the surface of the ramp, where $\mu$ is a material-specific friction coefficient and $N$ is the normal force
\begin{equation}\label{friction}
N = m (g \cos \theta + \ddot{h} + (L-h) \dot{\theta}^2).
\end{equation}
The force term \eqref{friction} results from balancing the force component normal to the ramp with gravity in the radial direction since $m(L-h)\dot{\theta}^2$ represents the centrifugal force. \\
Starting from an initial angle $\theta_0$ and zero initial velocity $\dot{\theta}_0=0$, the height $h$ is adjusted during a roll from the initial angle to the final angle $\theta^*$ at which all kinetic energy is consumed and the amplitude is maximal. Different height variations then lead to different maximal amplitudes $\theta^*$. We are interested in an optimal control function $\hat{h} = \hat{h}(t)$ such that
\begin{equation}\label{opt1}
\hat{h}(t) = \arg\max_{\{h(t)\}} \theta^*[h(t)],
\end{equation}
subject to the constraint $H_0 \leq h(t) \leq H_1$ and $|\ddot{h}(t)|\leq g$ for all times. Problem \eqref{opt1} of controlling a variable-length pendulum under different constraints and damping mechanics has been considered in, e.g.,~\cite{anderle2022controlling,lavrovskii1993optimal}.\\
%As the attained angle $\theta = \theta^*$ satisfying Equation \eqref{implicit1} is a functional of a variable length $r(u)$, the problem of maximizing the amplitude is expressed as \begin{equation}\label{opt1}\hat{h}(\theta) = \arg\max_{\{h(\psi)\}} \theta^*[h(\psi)],\end{equation}subject to the constraint $H_0 \leq h(\theta) \leq H_1$ and $|\ddot{h}|\leq g$. 
%\section{Angle Variables}
\begin{figure}[h]
\centering
\includegraphics[width=0.9\linewidth]{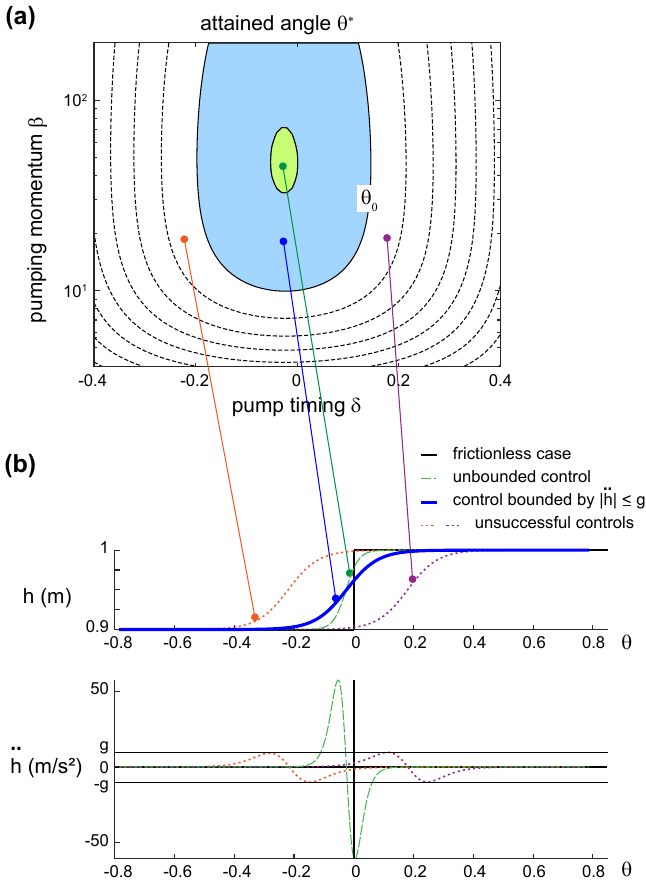}
\caption{\footnotesize{Pumping on a cylindrical ramp in the presence of kinetic friction. (a) The attained angle $\theta^*$ plotted against the pumping timing $\delta$ and pumping momentum $\beta$. (b) The optimal height $h$ and its acceleration $\ddot{h}$ in the frictionless case (solid black line); unbounded control in the presence of kinetic friction (green dashed line) and the optimal solution under the acceleration bounded by $|\ddot{h}| \le g$ (thick blue line). The optimal strategies for a cylindrical ramp are compared with unsuccessful controls where an achieved angle $\theta^*$ is smaller than the initial angle $|\theta_0|$ (brown and orange dotted lines). Parameters: $\theta_0=-\pi/4 \approx -0.79, L=3, H_0=0.9, H_1=1.0$ and $\mu=0.02$.}}
\label{curved}
\end{figure} 
To tackle the optimization problem \eqref{opt1} directly, we assume that the height is a monotonic function of time for a one-directional roll along the cylindrical ramp. Since the angle $\theta(t)$ along one roll is a monotonic function of time as well, we may rewrite the governing equation for the radius $r(\theta) = L-h(t)$ as a function of the angle, which will simplify the notation in the following. This allows us to derive an equation for the kinetic energy with respect to angular rotation $E_{\rm kin} =\dot{\theta}^2/2$ as a function of $\theta$. The analytical trick to write the kinetic energy and the radius as functions of the angle - the original dynamical variable - is reminiscent of the modern solution to the Brachistochrome problem \cite{erlichson1999johann} and similar problems in the calculus of variations.\\
To this end, we write $r'=dr/d\theta$ and note that $\dot{r}=r'\dot{\theta}$. By transforming the acceleration $\ddot{r}$ to angle-variables, $\ddot{r} = d/dt( r'\dot{\theta}) = \ddot{\theta} r' + \dot{\theta}^2 r''$, 
the equation of motion can be transformed into an equation for the kinetic energy $E_{\rm kin}$ with respect to $\theta$,
\begin{equation}\label{eqEkinfriction}
\begin{split}
r E_{\rm kin}'+ & 4r'E_{\rm kin}+g \sin \theta\\
&+ \mu (g \cos \theta - r' E_{\rm kin}'+ 2 (r-r'') E_{\rm kin}) =0. 
\end{split}
\end{equation}
The linear first-order differential equation \eqref{eqEkinfriction} can be integrated easily to 
\begin{equation}\label{eqEkinfrictionsol}
\begin{split}
E_{\rm kin}(\theta) & = -g \int_{\theta_0}^\theta \frac{\sin u + \mu \cos u}{r(u)-\mu r'(u)} \exp{\left( \int_{u}^\theta f(s) ds\right)} du,\\
f(s) & = -\frac{4r'(s)+ 2 \mu (r(s)-r''(s))}{r(s)-\mu r'(s)}.
\end{split}
\end{equation}
We remark that the influence of an air drag or an inertial drag proportional to the square of the velocity, $ F_{\rm air} = C_D |\mathbf{v}|\mathbf{v}$ can be treated analogously to kinetic friction and an explicit formula similar to \eqref{eqEkinfrictionsol} may be derived by explicitly solving a linear differential equation for the angular kinetic energy. The numerical results for the optimal control strategy, however, show little deviation as compared to pure kinetic friction and we focus on formula \eqref{eqEkinfrictionsol} henceforth. At the maximal amplitude $\theta^*$ along one monotonic roll, all kinetic energy is consumed and we have that $E_{\rm kin}(\theta^*)=0$. Setting $\mu = 0$ in \eqref{eqEkinfrictionsol} we immediately see that the optimal control strategy in the frictionless case is given by
\begin{equation}
\hat{r}_{\rm frictionless}(\theta) = \left\{
\begin{array}{cc}\label{frictionlessopt}
L-H_0 ,& (\theta_0 \le \theta \le 0),\\
L-H_1, & (0 < \theta),
\end{array}
\right.
\end{equation}
where we have assumed that the skateboarder starts his/her roll on the left side of the ramp ($\theta_0<0$). The maximal angle $\theta^*_{\rm frictionless}$ achieved by this optimal control can be calculated explicitly as well and is given by
\begin{equation}\label{frictionlesssol}
\cos \theta^*_{\rm frictionless} = 1- \left( \frac{L-H_0}{L-H_1}\right)^{3} (1-\cos{\theta_0}).
\end{equation}
Solutions \eqref{frictionlessopt} and \eqref{frictionlesssol} will serve as a benchmark for the full, physically meaningful optimal control with constraints.

%\section{Optimization Problem and Numerics}

\onecolumngrid

\begin{figure}[h]
\centering
\includegraphics[width=0.6\linewidth]{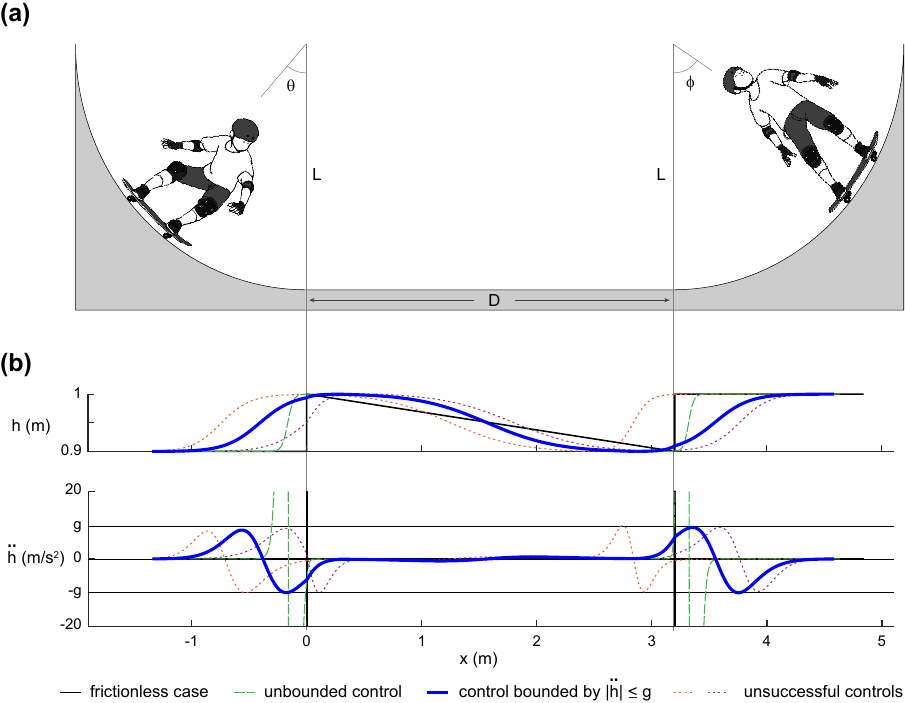}
\caption{\footnotesize{ (a) Simulations for the optimal pumping motion of a skateboarder on a half-pipe consisting of two cylindrical ramps connected by a flat zone in between. The radius of curvature of the cylindrical parts is $L=1.9$ m and the width of the flat zone is $D=3.2$ m, corresponding to the experimental conditions. (b) Height and acceleration variations for different optimal pumping strategies: optimal control for the frictionless case (black line), optimal unbounded control in the presence of kinetic friction (dashed green line), and optimal control under the acceleration bounded by $|\ddot{h}| \le g$ (blue thick line). The optimal control strategies are compared with unsuccessful pumping (brown and orange dotted lines), where the achieved maximal angle is smaller than the initial amplitude. }}
\label{halfpipe}
\end{figure}

\twocolumngrid

As potential pumping strategies, we examine a two-parameter family of logistic functions interpolating between maximal and minimal height,
\begin{equation}\label{frictionopt}
r(\theta)=L-H_0-\frac{H_1-H_0}{1+e^{-\beta(\theta-\delta)}},
\end{equation}
for the pumping momentum $\beta$ and the timing parameter $\delta$. For the numerical simulation, we choose a friction coefficient of $\mu = 0.02$, which is consistent with the experimentally obtained friction coefficient. We compare the optimal pumping strategies for the cylindrical parts of the half-pipe in Figure~\ref{curved} for the frictionless case, unbounded acceleration, acceleration bounded by $|\ddot{h}| \le g$, and unsuccessful pumping motions, where the achieved maximal angle $\theta^*$ is smaller than the initial amplitude $|\theta_0|$.

We continue with the analysis of the full half-pipe, starting with an initial position of the skateboarder at the left cylindrical part of the ramp $(\theta_0<0)$. To enter the flat zone linking the two cylindrical parts with the highest possible speed, we seek optimal parameters $\beta$ and $\delta$ from \eqref{frictionopt} such that $E_{\rm kin}$ is maximal at $\theta=0$ (entering the flat zone). During the motion in the flat zone, the height $h$ is decreased to $H_0$ continuously until leaving the flat zone again. Upon entering the right cylindrical part, we search for another set of parameters in \eqref{frictionopt} so that the final angle of a skateboarder $\theta^*$ is maximized. Similar to the cylindrical ramp, we note that in the frictionless case, the maximal amplitude can be calculated explicitly to 
\begin{equation}
\cos \theta^* = 1- \left( \frac{L-H_0}{L-H_1}\right)^{5} (1-\cos{\theta_0}),
\end{equation}
We remark that the maximal angle is larger than the one of the cylindrical ramp \eqref{frictionlesssol} since the skateboarder performed two pumping cycles in the half-pipe. Figure ~\ref{halfpipe} compares the optimal pumping strategies in the half-pipe for bounded acceleration (thick blue line), the frictionless case (solid black line), the case with friction but acceleration unbounded (dashed green line), and two examples of unsuccessful pumping (brown and orange dotted lines) where the achieved maximal angle is smaller than the initial amplitude.

%%%%%%%%%%%%%%%%%%%

\begin{figure}[h]
\centering
\includegraphics[width=0.8\linewidth]{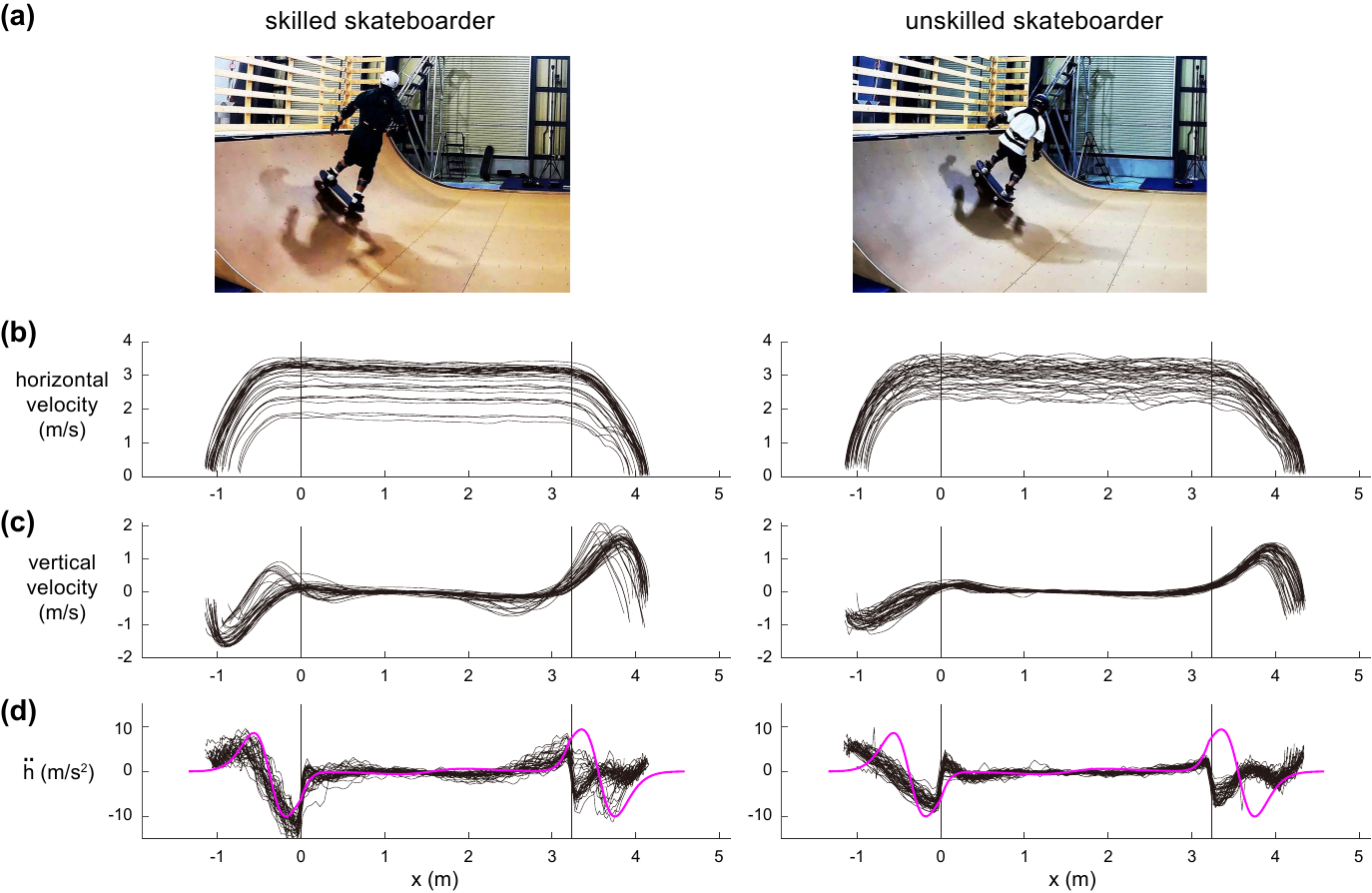}
\caption{\footnotesize{(a) Pumping on a half pipe performed by skilled and unskilled skateboarders. (b) and (c) The horizontal and vertical velocities of the center of mass. Each line represents a left-to-right motion segmented from the back-and-forth motion. (d) The acceleration of the height from the ramp surface, $\ddot{h}$. The magenta line shows the theoretical-numerical prediction of the optimal control strategy.}}
\label{experiment}
\end{figure}

%\section{Experiments}
For the experimental validation of the optimal control strategy, two skateboarders with different degrees of skill were asked to gain height as much as possible by adopting their most efficient pumping action. The skilled skateboarder has eleven years of experience, while the unskilled skateboarder has two years of experience. For each trial, they started from a position of rest and stopped their pumping when they reached the top of the ramp. The subject's center of mass was estimated automatically by the motion-capture software Xsens. The acceleration in the sagittal plane $\{a_x, a_y\}$ was translated into the height acceleration measured relative to the ramp surface $\ddot{h}$ using information about the center of mass of the subjects.\\
Figure~\ref{experiment}a shows snapshots of the skilled and unskilled skateboarders performing reciprocating during the experiment. Figures~\ref{experiment}b, c, and d show their horizontal velocity $v_x$ and vertical velocity $v_y$ as well as the acceleration $\ddot{h}$. Each black solid curve represents a left-to-right motion segmented from the back-and-forth motion.
\onecolumngrid

\twocolumngrid
Firstly, the friction coefficient $\mu$ is estimated from the decrease in horizontal velocity in Figure~\ref{experiment}b. With an average speed of $v \approx 3$ m/s, the skateboarder took $\Delta t \approx 1$ s to pass the flat zone of 3 m. The velocity decreased by $\Delta v \approx 0.2$ m/s during this passage, implying that the change in velocity $\Delta v / \Delta t = \mu g$ or $\mu \approx 0.02$. We also used this friction coefficient as an input parameter for the theoretical estimation described above.\\
Secondly, we have compared the estimated acceleration with the optimal acceleration obtained by our theory (Figure~\ref{halfpipe}b). The skilled skateboarder had a smaller root mean square error (3.9 m/s$^2$) with the theoretical optimal solution than the unskilled skateboarder (4.3 m/s$^2$). This indicates that our theoretical-numerical optimal pumping strategy is followed more closely by the skilled skateboarder as compared to the unskilled one.

%\section{Conclusions}

We conclude with a summary of the results. We modeled the pumping motion of a skateboarder in a half-pipe as a variable-length pendulum with kinetic friction in the cylindrical part of the ramp surface and solved the equation of motion explicitly for the kinetic energy in angular coordinates. This allows us to formulate an optimization problem for the maximal angle in one roll. The results were compared to an experimental study in which the pumping strategies of skilled and unskilled skateboarders trying to gain height were compared. The skilled skateboarder showed less deviation from the theoretically predicted optimal control strategy. 

By simplifying the highly complex and delicate movement of skateboarding into the low-degree-of-freedom dynamical problem, we were able to suggest potential improvements in the performance of an athlete. This type of simplified kinematic consideration could also be applied to investigate improvements in other sports, such as ski jumping.
\\

\begin{acknowledgements}
We thank Kazuo Tsuchiya, Eiji Uchibe, and Shin Takagi for constructive comments, and Kai Shinomoto for drawing the illustrations for Figure 3. D.E.C. and S.S. were supported by the New Energy and Industrial Technology Development Organization (NEDO). \\
\end{acknowledgements}

F.K. and S.K. contributed equally to this work. F.K. developed the analytical method and wrote the paper. S.K. performed the optimization and numerical analysis. D.E.C. performed the experiments. S.S. designed the study and drafted the manuscript.

\bibliographystyle{apsrev4-1}
\bibliography{references}

\end{document}